\input epsf
\input harvmac

\def\p{{\bf p}}

\def\r{{\bf r}}

\def\x{{\bf x}}

\def\frac#1#2{{#1\over #2}}

%
%
\ifx\epsfbox\UnDeFiNeD\message{(NO epsf.tex, FIGURES WILL BE IGNORED)}
\def\figin#1{\vskip2in}
\else\message{(FIGURES WILL BE INCLUDED)}\def\figin#1{#1}\fi
\def\ifig#1#2#3{\xdef#1{fig.~\the\figno}
\goodbreak\midinsert\figin{\centerline{#3}}%
\smallskip\centerline{\vbox{\baselineskip12pt
\advance\hsize by -1truein\noindent\footnotefont{\bf Fig.~\the\figno:}
#2}}
\bigskip\endinsert\global\advance\figno by1}

\def\ifigure#1#2#3#4{
\midinsert
\vbox to #4truein{\ifx\figflag\figI
\vfil\centerline{\epsfysize=#4truein\epsfbox{#3}}\fi}
\narrower\narrower\noindent{\footnotefont
{\bf #1:}  #2\par}
\endinsert
}

\Title{\vbox{\baselineskip12pt\hbox{Univ.Amsterdam, November, 1997}
		\hbox{cond-mat/9805010}}}
{\vbox{\centerline{Scattering from objects}
	\vskip2pt\centerline{immersed in a diffusive medium.}}}

\centerline{David~Lancaster\footnote{$^{(a)}$}
  {\tt(djl@ecs.soton.ac.uk)}
	and Theo~Nieuwenhuizen\footnote{$^{(b)}$}
  {\tt(nieuwenh@phys.uva.nl)}}

\bigskip
\centerline{Van der Waals--Zeeman Instituut,}
\centerline{University of Amsterdam,}
\centerline{Valkenierstraat 65,}
\centerline{1018XE Amsterdam.}

\vskip .3in
{\bf Abstract}: 
The disturbance of the transmission of light through a diffusive 
medium due to an object hidden in it can be expressed in terms of 
an effective  charge  and dipole moment.
In the mesoscopic regime, beyond the diffusion approximation,
we calculate this effective charge and dipole moment.
Various objects are considered: a single point scatterer,
a localised density of such scatterers and a small
sphere.

\Date{11/97} 

\lref\OuNi{P.N.~den~Outer, Th.M.~Nieuwenhuizen and Ad~Lagendijk,
	J.Opt.Soc.Am.A {\bf 10}, 1209 (1993).}
\lref\Phil{H.M.J.~Boots, J.H.M.~Neijzen, F.A.M.A.~Paulissen
	M.B.~van der Mark and H.J.~Cornelissen,  
	Liquid Crystals {\bf 22}, 255-264 (1997).}
\lref\LuNi{Th.M.~Nieuwenhuizen and J.M.~Luck,  Phys.Rev. {\bf E 48}, 569
	(1993).} 
\lref\Dutch{Th.M.~Nieuwenhuizen, Lecture notes (in Dutch)
	(University of Amsterdam, 1993)
	\semi
	M.C.W.~van~Rossum and Th.M.~Nieuwenhuizen, 
	submitted to Rev. Mod. Phys.} 
\lref\Chan{S.~Chandrasekar, {\it Radiative Transfer}, (Dover, New York,
1960).}
\lref\Roa{M.V.~Rossum, PhD thesis, (Amsterdam, 1996).} 
\lref\NLvT{Th.M. Nieuwenhuizen, A. Lagendijk, and B.A. van Tiggelen,
Physics Lett. A {\bf 169} (1992) 191.}
\lref\Mierefs{J.D.~Jackson, {\it Classical Electrodynamics}, Wiley, New York
	(1975)
	\semi
	J.J.~Bowman, T.B.A.~Senior and P.L.E.~Uslenghi (eds),
	{\it Electromagnetic and Acoustic Scattering by Simple Shapes},
	North Holland, Amsterdam (1969).}
\lref\BeFe{Berkovits and Feng [75], Phys.Rev. {\bf B 22}, 5553 (1980).} 
\lref\LuN{J.M.~Luck and Th.M.~Nieuwenhuizen, to appear.}
\lref\Paas{ J.C.J. Paasschens and G.W. 't Hooft, J. Opt. Soc. A., submitted;
 J.C.J. Paasschens, PhD thesis (Leiden, 1997).}

\newsec{Introduction}

The reconstruction of the properties of an object from data obtained
in a scattering experiment is 
a problem that arises in all branches of physics.
Here we consider the scattering of light from an object placed 
inside a diffuse medium. This poses problems
because of the many further scattering events between
object and observer; in everyday language, ``the object cannot be seen''. 
However, with the help of sufficiently intense light sources, and careful
and sensitive experimental techniques the small perturbations
on the diffuse light due to the embedded object become detectable.
For example experiments have been performed by den~Outer, Nieuwenhuizen 
and Lagendijk \OuNi.
The technological applications to non-invasive methods
of detecting tumors in human flesh are active areas of research.

The significant feature of this work is that we go beyond the 
diffusion approximation discussed in \OuNi\
and make analytic calculations based on  
mesoscopic theory. This means that although the scattering 
length is much longer than the wavelength of the light 
($kl \gg 1$), we do not require that the relative
size of the objects be much larger than the scattering length.
Furthermore, we are able to impose boundary conditions
based on the microscopic theory, which cannot be done in
the diffusion approximation. We start with the simplest case,
namely a point object.
Our main result is for a small spherical object and
we expect this result to be useful in
providing an independent check of the numerical and other 
approximate approaches sometimes used in this field \Phil.

We make several simplifying assumptions: most importantly 
we only consider scalar waves and always assume isotropic scattering.
Our results are presented in the form of the effective charge 
and dipole moment of the multipole expansion of the solutions
of the diffusion equation in the far field region.
This allows the results to be adapted to whatever detection
arrangement is desired.

The paper starts with a description of the salient points of 
the mesoscopic multiple scattering formalism forming the basis
of our analysis. The following sections, 3,4,5 
consider a series of progressively more realistic objects:
a point scatterer, a collection of such scatterers and a
small sphere subject to a variety of boundary conditions. 
A short conclusion summarises our results.


\newsec{Mesoscopic Scattering Formalism}

We base our approach on the Schwarzchild Milne (SM) equation of radiation
transfer theory. This equation may be obtained by integrating
the full radiation transfer equation \Chan\ but here we prefer to 
regard it as a Bethe Saltpeter--like integral
equation for the diffuse intensity
$J({\bf r})$. We write it in the general form,
\eqn\SMfreedef{
J({\bf r}) = S({\bf r}) + \int d^3r' {\cal M}({\bf r},{\bf r}') J({\bf r}').
}
Where the source $S$  and the kernel ${\cal M}$ take
different forms depending on the geometry of the problem.
In the ladder approximation 
the kernel is given by the square of the amplitude
green functions: 
${\cal M}  = {4\pi\over l} |G({\bf r},{\bf r}')|^2$.
A uniform density of scatterers leads to
an amplitude green function $G_0(r) = e^{i(k+i/2l)r}/4\pi r$,
where the scattering length $l$ is related to the $t$-matrices of the
individual scatterers and their density ($n$) by $l = 4\pi/nt\bar t$.
It is then simple to
see that ${\cal M}_0 =  e^{-r/l}/4\pi lr^2$
(the SM equation is often stated for a slab geometry, and
the relevant kernel is then obtained from ${\cal M}_0$ by integrating
over perpendicular momenta).
The Fourier transform of this kernel is 
$\tilde {\cal M}_0(q) = {\tan^{-1}(ql)/ ql}$
($= A_1(ql)$ in the notation described in appendix C),
so the green function ${\cal G}_0$ for intensity 
propagation in the SM equation is
${\cal G}_0(q) = 1/(1-A_1(ql))$.
A fuller discussion of notation is provided in appendices A and B,
and a general review of the formalism is given in \Dutch.

In order to find the effect of an object embedded in the
multiple scattering medium we require an
expression for the appropriate SM kernel, which we base on a
perturbative expansion of the amplitude green function.
Firstly however, we consider exactly what quantity should
be calculated to compare with measurements.

\subsec{Measurement}

We assume that system boundaries are distant from the 
localised region of the object. In this far field 
region, the diffuse intensity $J({\bf r})$ obeys the diffusion
equation which is much simpler than the full SM equation.
Since we are dealing with static 
quantities, the diffusion equation reduces to the Poisson
equation.
\eqn\ab{
\nabla^2 J({\bf r})=0,
}
In the absence of the object, the solution of
equation \ab, subject to appropriate boundary conditions,
is  $J_0({\bf r})$.
The total intensity in the presence of the object then reads
\eqn\ac{
J({\bf r})=J_0({\bf r})+\delta J({\bf r}).
}
The goal of the present work is to characterise
the disturbance $\delta J({\bf r})$ of the diffuse intensity,
representing the diffuse image of the immersed object.
We make use of the fact that
the Poisson equation \ab, allows electrostatic analogies,
and represent the disturbance $\delta J({\bf r})$
in terms of a multipole series.
It turns out that only the first two of them,
namely the {\it charge} $q$ and the {\it dipole} $\p$,
play a role, while higher-order multipoles are negligible.
Assume for a while that the turbid medium is infinite.
The disturbance of the intensity far enough
from an object placed at ${\bf r}_0$ assumes the form
\eqn\diffsol{
\delta J({\bf r})=
\frac{q}{|{\bf r}-{\bf r}_0|}-\frac{{\bf p}\cdot({\bf r}-{\bf r}_0)}
{|{\bf r}-{\bf r}_0|^3}.
}
This expression, which is a solution of the diffusion equation \ab,
is also a solution of the full SM equation provided
the distance from the object
is large with respect to the mean free path {\it l}.

For a symmetric object, the linearity and isotropy of the problem
imply that the charge and dipole have the form
\eqn\ae{
q=-QJ_0({\bf r}_0),\quad\p=-P\nabla J_0({\bf r}_0).
}
We follow the notation  proposed by J.M.~Luck, and call
$Q$ the {\it capacitance} of the object,
and $P$ the {\it polarizability} \LuN.
These two numbers are intrinsic to the embedded object.

The capacitance $Q$ is non-zero (and in fact positive)
only if the sphere absorbs light.
For a non-absorbing object,
the leading contribution is that of the induced dipole $\p$.
The corresponding polarizability $P$ can be either positive or negative,
depending on the nature of the object.
Notice that $Q$ has the dimension of a length,
while $P$ has the dimension of a volume.

For real measurements
we should bear in mind that the diffusion
approximation will break down in the vicinity of the
measurement apparatus. For the particular case of a thick
slab geometry, which is often encountered experimentally \OuNi,
we now recall the derivation of the diffuse image of a small object
to show how the capacitance and polarizability can be measured,

The sample is a thick slab $(0<x<L)$, with $L\gg{\it l}$,
assumed to be infinite in the two transverse directions.
We set $\rho=(y^2+z^2)^{1/2}$.
The left side of the sample $(x=0)$ is subject to an incident plane wave,
so that the solution of eq. \ab$\,$ when the immersed object is absent reads
\eqn\af{
J_0(\r)=J\left(1-\frac{x}{L}\right). }
The intensity $T(y,z)$ transmitted through the sample,
and emitted on the right side $(x=L)$ at the point $(y,z)$,
is proportional to the derivative $\partial J(x=L,y,z)/\partial x$.
When there is no immersed object, eq. \af $\,$ yields a uniform transmission
$T_0=K{\it l}/L$.
The prefactor $K$ describes the angular dependence of the 
transmitted intensity.
It also depends on microscopic characteristics of the medium,
such as any anisotropy of the scatterers.

Let the immersed object have
capacitance $Q$ and polarizability $P$, and be located at $x=x_0$, $y=z=0$.
Its charge and dipole then read
\eqn\ag{
q=-QJ\left(1-\frac{x_0}{L}\right),\quad\p=\frac{PJ}{L}\hat x,
}
with $\hat x$ being the unit vector of the positive $x$-axis.

Along the lines of ref. \OuNi,
the total intensity can be determined by summing the free-space
expression \diffsol   $\,$ 
over an infinite double array of images $(-\infty<n<+\infty)$.
We thus obtain
\eqn\ah{\eqalign{
J(\x)=J\left(1-\frac{x}{L}\right)&+q\sum_n\left(
\frac{1}{\big[(x-x_0+2nL)^2+\rho^2\big]^{1/2}}
-\frac{1}{\big[(x+x_0+2nL)^2+\rho^2\big]^{1/2}}\right)\cr
&-\frac{PJ}{L}\sum_n\left(\frac{x-x_0+2nL}
{\big[(x-x_0+2nL)^2+\rho^2\big]^{3/2}}
+\frac{x+x_0+2nL}{\big[(x+x_0+2nL)^2+\rho^2\big]^{3/2}}\right).\cr
}}

The transmitted intensity $T(y,z)=T(\rho)$ then reads
\eqn\ai{
\eqalign{
T(\rho)=T_0\Bigg(
1&-2Q(L-x_0)\sum_n\frac{L-x_0+2nL}{\big[(L-x_0+2nL)^2+\rho^2\big]^{3/2}}\cr
&-2P\sum_n\frac{2(L-x_0+2nL)^2-\rho^2}{\big[(L-x_0+2nL)^2+\rho^2\big]^{5/2}}
\Bigg).
}}
Such characteristic profiles have been observed experimentally
for the case of pencils (black cylinders) and  glass fibers 
(transparant cylinders with index mismatch) by den Outer et al.

In order to calculate the charge and dipole moment in terms of
microscopic properties of the object we relate \diffsol\
to the far field behaviour of the full SM equation with 
the same incoming intensity. 
At first order, an object placed in the medium
modifies the free kernel of the SM equation by adding a vertex term
${\cal M} = {\cal M}_0 + V$,  which
for small momenta we can write\foot{We define the Fourier transform
so that the first momentum is incoming and the second outgoing.} as,
\eqn\vertex{
V({\bf p},{\bf p}') = 
{4\pi l^2\over 3}\left( 
-Q + P{\bf p}.{\bf p}'\right).
}
The parameters $Q$ and $P$ are precisely the ones that appear in 
the multipole expansion \diffsol. To see this, insert the 
expansion of the kernel, ${\cal M} = {\cal M}_0 + V$, into
the the SM equation and solve perturbatively for the far field
intensity $J({\bf r})$. This can be done using 
the diffusion limit of the free intensity green function,
$\tilde{\cal G}_0(r) \propto 1/r$.
We emphasise the pertubative nature of this general result,
and indeed it fails in the case of a large spherical object.
This will be the reason for later restricting our attention to
a small sphere.

In the rest of the paper we will define the vertex more precisely
and develop a diagrammatic perturbation expansion of the SM equation
to calculate it both at leading and higher order. Only the small
momentum behavior of the vertex is important for our goal
of calculating $Q$ and $P$, and we shall find that in this
case the higher order corrections can sometimes be resummed.


\newsec{Additional Point-Like Scatterer}

The effect of an object consisting of a single  
point-like scatterer is a worthwhile
exercise because it allows us to demonstrate the method 
in a simple setting. 
Although the higher order terms can be formally calculated,
they are not reliable since we encounter
ill-defined expressions arising from the point-like limit
which take us out of the realm of validity of the mesoscopic
theory.

\subsec{Leading Order Calculation}

Starting from the amplitude green function for a single extra
scatterer with $t$-matrix $t_e$, at the origin,
\eqn\deltaG{
G({\bf r},{\bf r}') = G_0(|{\bf r}-{\bf r}'|) 
+ G_0({\bf r}) t_e  G_0({\bf r}').
}
The full SM kernel is given by,
\eqn\deltaM{
{\cal M}({\bf r},{\bf r}') = {\cal M}_0 + \delta {\cal M} 
= {4\pi\over l} | G |^2 = 
 {4\pi\over l}\vert G_0(|{\bf r}-{\bf r}'|)
 +  G_0({\bf r}) t_e  G_0({\bf r}')\vert^2.
}
\ifig\vertexone{
Diagrams contributing to the bare vertex $V_1$.}
{
\epsfysize=2.5cm 
\epsfbox{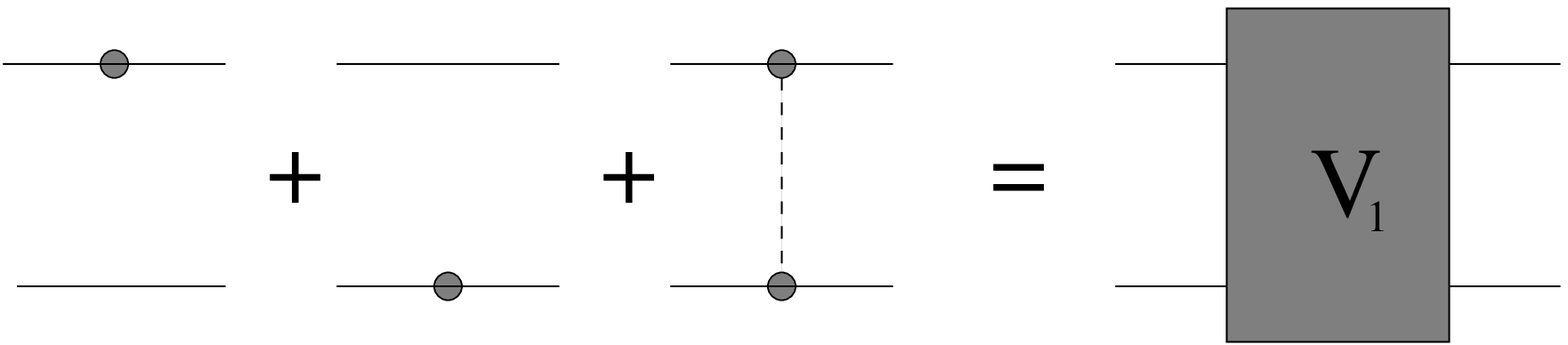}}
The diagrammatic interpretation is obvious and the 
vertex \vertex\ contains the terms shown in figure 1.
All terms will be the same order due to the optical theorem.
If we evaluate the diagrams in momentum space we may use the
formulae in appendix B to obtain the mesoscopic limit,
\eqn\ExtrascatA{
V_1({\bf p},{\bf p}')  = \delta \tilde{\cal M}({\bf p},{\bf p}') =
l
\Biggl( {t_e \bar t_e \over 4\pi} A_1(pl)A_1(p'l)
+ {i(t_e - \bar t_e) \over 2 k} 
A_2({\bf p}l,{\bf p}'l)
\Biggr).
}

Now we use the optical theorem as applied to the additional
scatterer. We allow this extra scatterer to have a complex
refractive index and introduce the albedo $a_e$ 
which appears in the optical theorem:
\eqn\Optical{
\sigma_{el} =
{t_e \bar t_e \over 4\pi}
= a_e {\Im m\, t_e \over k}
}
Where $\sigma_{el}$ is the elastic contribution to the cross section for the
additional scatterer.

The final expression for the bare vertex is,
\eqn\Extrascatbare{
V_1({\bf p},{\bf p}')  =
l\sigma_{el}
\biggl( A_1(p l)A_1(p'l)
-{1\over a_e} A_2({\bf p}l,{\bf p}'l) 
\biggr)
}
By using the small momentum expansions of the $A_i$ functions 
given in appendix C, we are able to identify the charge and
dipole moment defined by equation \vertex,
\eqn\QandPone{\eqalign{
Q_1 &= {\sigma_{el}\over l} {3\over 4\pi} (1/a_e - 1)
= {3\over 4\pi}{\sigma_{abs}\over l} \cr
P_1 &= {\sigma_{el}l} {1\over 4\pi} (1/a_e) 
={1\over 4\pi}{\sigma_{tot}l}\cr
}}
For albedo $a_e < 1$, there is absorption and the charge is positive.

\subsec{Second Order Correction}

This expression for the vertex, $V_1 = \delta {\cal M}$, is only the first
term in the perturbative expansion of the SM equation. Higher 
order terms contain these vertices connected by intensity propagators
in the form $V_n = V_1{\cal G}_0V_1\dots V_1{\cal G}_0V_1$. 
In general these vertices are not easy to evaluate because the intermediate
loop integrals do not give rise to closed expressions.
However, for the multipole expansion we only require the  
behaviour for small external momentum
and can therefore approximate the first and last $V_1({\bf p},{\bf p}')$.
In this case, the 
angular integration over the intermediate momenta
simplifies the expressions remarkably 
so that the effect propagates though the 
series making it valid for each loop as shown below.

For small incoming momentum, $p$, we can expand the bare vertex 
using the expansion of $A_2$ in appendix C. This leaves 
$V_1({\bf p},{\bf p}')$ written in terms of 
$A_1(p')$ and explicit factors of momenta:
$p^2$, $p'^2$
and ${\bf p}.{\bf p}'$.
Let us work to first order in $p$ in which case,
\eqn\Extrascatexp{
V_1({\bf p},{\bf p}')
\buildrel {p \rightarrow 0} \over \longrightarrow
-{4 \pi l^2 \over 3}
Q_1 A_1(p'l) 
+ {\bf p}.{\bf p}'4\pi l^2 P_1  
A_3(p'l) 
}
where $Q_1$ and $P_1$ are given above in \QandPone\ and we have employed
the useful combination $A_3(q)=(1- A_1(q))/q^2$.

\ifig\vertexone{
The second order vertex $V_2$, is composed of two first order vertices.}
{
\epsfysize=2.5cm 
\epsfbox{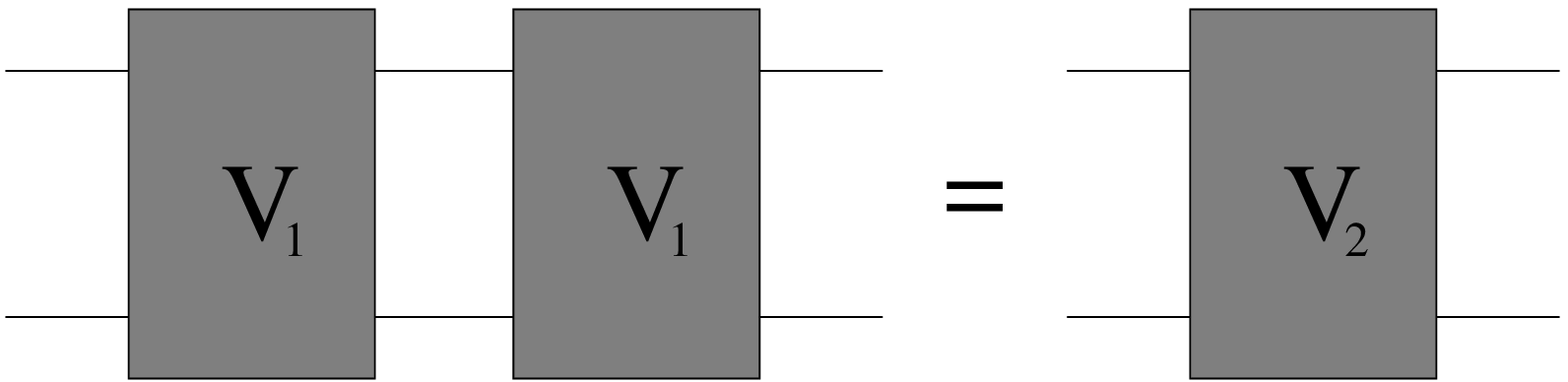}}
Now consider the second order contribution shown in figure 2,
\eqn\Firstdiag{
V_2({\bf p},{\bf p}')
=\int \! {d^3q\over (2\pi)^3} \ 
V_1({\bf p},{\bf q})
{\cal G}_0(q)
V_1({\bf q},{\bf p}')
}
where ${\cal G}_0$ is the free propagator as dicussed in section 2.
Using the approximate expression \Extrascatexp\ in the first
factor of $V_1$ the integrals become tractable. 
Performing the angular part of the integral using the
formulae given in appendix C we find that
the general form of $V$ is preserved,
\eqn\Firstorder{
V_2({\bf p},{\bf p}')=
-{4 \pi l^2 \over 3}
Q_2 A_1(p'l) 
+ {\bf p}.{\bf p}'4\pi l^2 P_2  
A_3(p'l) 
}
where the explicit expressions for $Q_2$ and $P_2$ are,
\eqn\Qone{\eqalign{
Q_2 &= 
  {\sigma_{el}\over l} \left({1\over a_e}-1\right)B_1 Q_1
= -{4\pi\over 3}B_1 Q_1^2\cr
P_2 &= 
  -{\sigma_{el}} \left({1\over a_e l^2}\right) B_2 P_1 
= -4\pi B_2 P_1^2\cr}
}
The remaining integrations over the magnitude of the momenta are
contained in the dimensionless factors $B_1$ and $B_2$, where
in the second expressions we have used the explicit form of 
${\cal G}_0(q)$.
\eqn\FirstQP{\eqalign{
B_1 &= {l^2\over 2\pi^2}\int dq\ q^2 A_1^2(ql){\cal G}_0(q)
= {1\over 2\pi^2 l}\int_0^{\Lambda l} dq\ q^2 {A_1^2(q)\over 
(1-A_1(q))}\cr
B_2 &= {1\over 2\pi^2}\int dq \  q^2  A_3^2(ql){\cal G}_0(q)
= {1\over 2\pi^2 l^3}\int_0^{\Lambda l} dq \ (1-A_1(q))\cr
}}
Here we encounter a problem due to the point-like limit. The
integrals diverge linearly, so we have introduced a high momentum regulator 
$\Lambda$ which we expect to be of the order of the inverse
size of the scatterer, $\Lambda\sim 1/R$. For the point-like
analysis above to be valid, the size of the extra scatterer
must be much less than the wavelength, $R << \lambda$.
This means that very high momenta, 
up to the inverse scale of the scatterer, are encountered
in the integral and in this range the mesoscopic approximation
fails. The expressions for the second order terms
are therefore formal. A similar divergence arises for the $t$-matrix of a 
point scatterer \NLvT, where it was parametrised as 
a linear divergence depending on some property of the scatterer.
The problem of divergencies will not appear in the following sections where
we will consider extended objects.

\subsec{Higher Order Corrections}

Bearing in mind the caveats above, we proceed formally to 
illustrate how the higher order effects may be calculated.
Since the form of the vertex remains the same as additional
bubbles are added, it is clear that the full series can be summed
to obtain an expression for the full $V$. 
The explicit expressions for $Q$ and $P$ are then,
\eqn\FullQP{\eqalign{
Q &= {Q_1\over 1 + {\sigma_{el}\over l}(1/a_e-1) B_1}
= {Q_1\over 1 + {4\pi\over 3}Q_1 B_1}
\cr
P &= {P_1\over 1 + {3\sigma_{el}l}(1/a_e) B_2}
= {P_1\over 1 + 4\pi P_1 B_2}
\cr
}}
It is interesting to note that if the capacitance $Q$ is negative 
(corresponding to amplification rather than absorption)
it could be substantially
enhanced by these higher order effects.

\subsec{Maximally Crossed Diagrams}

Because we are dealing with the peculiar situation of a point-like 
object another issue becomes important, which although academic
is interesting to discuss.
Crossed diagrams generally play a sub-leading role in mesoscopic
calculations, for example in the ladder propagator 
it can be seen that crossed diagrams contribute
at higher order in $1/kl$.
This conclusion must be reexamined for the point object and
diagrams in real space show that the
crossed diagrams can contribute at the same order as uncrossed ones. 
This can be understood as being due to the phase cancelling in the same way 
as it does for enhanced backscattering.
These contributions only occur when maximally crossed series are
inserted between two additional scatterers on the same line
(there can be an extra scatter on the other line or not).
In this case, each diagram contributes exactly as for the ladder sum.
The previous  calculation must be generalised
so that crossed diagrams are not included between additional
scatterers on different lines.
This combinatoric problem may be solved using a matrix formalism
described in Appendix D
and the final result in the mesoscopic limit is that the 
charge and dipole in that case are as above but $B$ gains a
factor of 2.
%
%

These calculations illustrate how the full sum can be
calculated from an infinite series of diagrams.
Unfortunately the results are flawed because the divergences
due to the point nature of the object take the calculation
out of the realm of validity of the mesoscopic approximation.
In the next sections we will consider better defined systems.


\newsec{Many Extra Scatterers}

We now consider an object consisting of a collection of extra 
point-like scatterers localised in some region of size $R$.
We first develop perturbation theory for the amplitude green functions
which we then use in the kernel of the SM equation.

\subsec{Perturbative expansion}

Consider extra scatterers placed in a localised region with density
profile $n_e({\bf r})$. The perturbative expansion of the 
amplitude green function in terms of the individual extra scatterer
t-matrix $t_e$, resembles the expression \deltaG,
\eqn\pertamp{
 G({\bf r},{\bf r}')  = 
 G_0(|{\bf r}-{\bf r}'|)  
+t_e\int d^3r'' G_0(|{\bf r}+{\bf r}''|)  n_e(r'')
 G_0(|{\bf r}''-{\bf r}'|)  +\dots
}
Where, as usual, $G_0$ is the green function in the presence
of a constant density of background scatterers.
In momentum space we have,
\eqn\pertampp{\eqalign{
 \tilde G&({\bf p},{\bf p}')  = 
 \delta({\bf p}-{\bf p}') G_0({\bf p})
+\tilde n_e({\bf p}-{\bf p}') t_e G_0({\bf p}) G_0({\bf p}') \cr
&+ \int {d^3q\over (2\pi)^3} 
\tilde n_e({\bf p}-{\bf q})\tilde n_e({\bf q}-{\bf p}') t_e^2
 G_0({\bf p}) G_0({\bf q}) G_0({\bf p}')+\dots\cr
}
}
We use the above expression to find a form for the SM kernel.
Here we must be careful to include an extra term in the definition of the
kernel because the
amplitude lines may end on an extra rather than background 
scatterer; to the factor $4\pi/l$ interpreted as
$n|t|^2$, we must add the contribution of the extra scatterers
$n_e(r)|t_e|^2$.
\eqn\mvBSkernel{\eqalign{
{\cal M}({\bf r},{\bf r}') &= \left({4\pi \over l} + n_e(r)|t_e|^2\right) 
| G ({\bf r},{\bf r}')|^2 \cr
\tilde{\cal M}({\bf p},{\bf p}') &= 
{4\pi \over l} 
\int {d^3q\over (2\pi)^3}{d^3q'\over (2\pi)^3}
 \bar G ({\bf q}+{\bf p},{\bf q}'+{\bf p}') G ({\bf q},{\bf q}')\cr
& + |t_e|^2
\int {d^3q\over (2\pi)^3}{d^3q'\over (2\pi)^3}{d^3q''\over (2\pi)^3}
\tilde n_e({\bf q}'')
 \bar G ({\bf q}+{\bf p},{\bf q}'+{\bf q}''+{\bf p}') 
G ({\bf q},{\bf q}')\cr}
}

In this problem, the vertex that relates the mesoscopic SM calculation to
the multipole expansion \diffsol\ is not simply given by $V = \delta {\cal M}$
as it was for the single point object.
In the far-field region where the identification is made, there are no
extra scatterers, so we must insist that the vertex only contain
diagrams which  end on a background rather than extra scatterer.
This prescription is most easily viewed diagrammatically rather
than in terms of the formal perturbation expansion of the SM equation
and its effect is also to remove the asymmetry apparent in the kernel itself.

Another difference with the point scatterer case is that
the perturbation expansion is organised in powers of the extra
density factor $n_e(r)$, and we are unable to relate different orders with the
Optical theorem. The  additional term included in \mvBSkernel\
saves the situation.
Inserting the expansion of the green function and 
identifying the first order term  ${\cal M}_1$, 
the additional term also contributes as the last term below.
\eqn\Vone{\eqalign{
\tilde{\cal M}_1({\bf p},{\bf p}') &= {4\pi \over l} 
\int {d^3q\over (2\pi)^3}{d^3q'\over (2\pi)^3}\cr
&t_e \tilde n_e(q-q') 
 G_0({\bf q}) G_0({\bf q}') \bar G_0({\bf q}+{\bf p})
\delta({\bf q}+{\bf p}-{\bf q}'-{\bf p}')\quad +{\rm c.c.}\cr
&+ |t_e|^2 \tilde n_e(p-p')
\int {d^3q\over (2\pi)^3}
 G_0({\bf q}) \bar G_0({\bf q}+{\bf p})\cr }
}
This expression can be simplified using the functions $A_i$
defined in appendix B and C. 
In order to obtain $V_1$ from ${\cal M}_1$, we
make a small modification to the second term in order that it
end on a background scatterer.
\eqn\VoneA{\eqalign{
V_1({\bf p},{\bf p}') &= \left({4\pi \over l}it_e 
\tilde n_e({\bf p}-{\bf p}')
{l^2\over 8\pi k} A_2({\bf p}l,{\bf p}'l')+ {\rm c.c.}\right)
+ |t_e|^2 \tilde n_e({\bf p}-{\bf p}'){l\over 4\pi} A_1(pl) A_1(p'l')\cr
&= \tilde n_e({\bf p}-{\bf p}')l\sigma_{el}
\left(A_1(pl)A_1(p'l') -{1\over a_e}A_2({\bf p}l,{\bf p}'l')\right)\cr}
}
The second line follows from the optical theorem applied to the
individual scatterers, as in the point scatterer case. 
We can immediately identify the charge and dipole at this order by
taking the momenta small and using the definition \vertex.
\eqn\Qone{\eqalign{
Q_1 &= 
  {\sigma_{el}\over l} {3\over 4\pi} N_e\left(1/a_e-1\right)
  = N_eQ_1^{point\ object}\cr
P_1 &= 
  {\sigma_{el}l}\, {1\over 4\pi} N_e\left(1/a_e 
  - 6{R^2\over l^2}(1/a_e-1)\right) 
  = N_eP_1^{point\ object}-2N_e{R^2}Q_1^{point\ object}\cr}
}
Where $N_e = \tilde n_e(0)$ is the total number of extra scatterers.
The second term in $P_1$ comes from expanding
the density form factor for small momentum as $\tilde n_e(p) \approx
N_e(1 - p^2R^2)$, and $R$ is a measure of the size of the object,
$R^2 = {4\pi\over 3N_e}\int r^4 n_e(r) dr$.
A more precise expression can be obtained if we assume a particular
density profile, for example a Gaussian form.
We have assumed a spherically symmetric density perturbation
so $\tilde n_e(p)$ is real.
The similarity with the single scatterer result \QandPone\ is clear; 
the charge merely gets multiplied by the number of scatterers present 
whereas the
dipole moment also picks up a term that depends on the 
size of the object.

\subsec{Second Order}

Two types of term occur at second order, those from
second order in the expansion of the amplitude green function,
and those from first order in the expansion of the amplitude green function
but second order in the expansion of the SM equation.

The first type of term can be neglected because for example
one contribution to the charge is,
\eqn\Qonee{\eqalign{
{4\pi \over l} |t_e|^2
\int {d^3q\over (2\pi)^3}{d^3q'\over (2\pi)^3}
|\tilde n_e({\bf q}-{\bf q}')|^2 
| G_0({\bf q})|^2| G_0({\bf q}')|^2\cr}
}
which corresponds to processes in which the amplitudes scatter
off different extra scatterers so the phase cancelation
kills them. This can be seen explicitly for a Gaussian profile,
where the contribution will be of order 
$\sigma_e/l^2 \times N_e^2/k^2R^2$. 
By comparing with the main second order contributions below
we obtain a condition on the size of object for this analysis 
to remain valid, 
$k^2R^2(1-1/a_e)B_1Q_1^{point\ object} \gg 1$. 

We therefore only consider the term with the form
$V_2 = V_1{\cal G}V_1$. Note that although this is clear
diagrammatically, at the level of the formal expansion
the correction we made to obtain $V_1$ from  ${\cal M}_1$ 
appears in $V_1{\cal G}V_1$ in just the right way to
count the relevant terms in ${\cal M}_2$, which need not be 
evaluated separately.
We attempt to use the same trick as for the point scatterer and
therefore expand the bare vertex for small incoming momentum $p$,
\eqn\Extrascatexpp{
V_1({\bf p},{\bf p}')
\buildrel {p \rightarrow 0} \over \longrightarrow
\tilde n_e ({\bf p}-{\bf p}')
\left(-{4 \pi l^2 \over 3}
Q_1^{point\ object} A_1(p'l) 
+ {\bf p}.{\bf p}'4\pi l^2 P_1^{point\ object} 
A_3(p'l)\right) 
}

Now consider the first diagram,
\eqn\Firstdiag{
V_2({\bf p},{\bf p}')
=\int \! {d^3q\over (2\pi)^3} \ 
V_1({\bf p},{\bf q})
{\cal G}(q)
V_1({\bf q},{\bf p}')
}
Because of the presence of the density form factor, the angular integrals
that kept the form of $V$ the same at higher order and therefore
allowed us to perform the full sum in the point scatterer case
cannot be done. 
Let us instead do the second order calculation by setting {\it both} the 
external momenta small.
The final expressions for $Q_2$ and $P_2$ are then,
\eqn\Qtwoe{\eqalign{
Q_2 &= 
  {\sigma_e\over l} N_e \left(1 -{1\over a_e}\right)B_1 Q_1\cr
P_2 &= 
  {\sigma_e l} N_e {1\over a_e} B_2 P_1
  -2{R^2} Q_1\cr}
}
Where the factors $B$ are now well-defined and do not need regulating
because the density
acts as a smooth cutoff at momenta of order $1/R$. In this regime
the mesoscopic approximation is still valid so the approach is consistent.
\eqn\FirstQP{\eqalign{
B_1 &= {1\over 2\pi^2N_e^2 l}\int_0^{\infty} dq\ q^2 
\tilde n_e^2(q/l) {A_1^2(q)\over 
(1-A_1(q))}\cr
B_2 &= {1\over 2\pi^2N_e^2 l^3}\int_0^{\infty} dq \ \tilde n_e^2(q/l)
(1-A_1(q))\cr
}}

This resolves the difficulties encountered in the point object 
case, but the discussion above shows that we cannot write
simple expressions for terms higher than second order.
Crossed diagrams are also suppressed in this case.


\newsec{Spherical Object}

We consider the simplest example of an extended object immersed in
the multiple scattering background: a sphere of radius $R$ ($kR \gg 1$)
subject to a variety of boundary conditions. The analogous problem 
in a free background was much studied in the early part of
this century \Mierefs, and some of the results relevant to the
present problem are collected in appendix E. 

The general result relating the effective charge to the fourier
transformed vertex is pertubative about the free theory, but
depending on the boundary conditions, the interior region
of a finite size object may not be susceptible to such treatment.
For example, for reflecting boundary conditions
the full green function in the interior vanishes, which 
cannot be regarded as a small correction.
A different approximation scheme, or an object with similar
refractive index to the medium could be considered, but here
we shall proceed by assuming that the sphere is small: $R \ll l$.
In this case it can be argued that the correction from the 
interior of the sphere is smaller than the perturbative term we shall
calculate. This limit is the opposite of the limit in
which diffusion theory is valid. 
An alternative approach to this problem is to use 
radiative transfer theory and boundary
conditions for the intensity field~\LuN.

\subsec{Leading Order}

Employing spherical polar coordinates
we write the amplitude Green function in terms of spherical harmonics
and spherical Bessel functions,
\eqn\sphereg{\eqalign{
G({\bf r},{\bf r}') &= \sum_{lm} g_l(r,r') Y_{lm}(\theta,\phi)
\bar Y_{lm}(\theta',\phi')\cr
g_l(r,r') &= g^{(0)}_l(r,r')+a_lg^{(1)}_l(r,r')
=ik\bigl( j_l(kr_<) + a_l h^{(1)}_l(kr_<)\bigr)\,h^{(1)}_l(kr_>)\cr
}}
Where the notation $r_>,r_<$ indicates the larger or smaller of the two
radii $r,r'$. 
As usual, $k$ has an imaginary part to take account of multiple scattering,
but on the surface of the sphere where the boundary conditions 
determining the coefficients $a_l$ are imposed,
the value of $k$ must be reconsidered.
In a fully self consistent approach, the imaginary part of $k$
is fixed by the return green function $G({\bf r},{\bf r})$ \Dutch,
and can vary in space. This variation is only significant close to
the surface of the extended object and in the mesoscopic approximation
it can be ignored\foot{In a one-dimensional version of the problem
this can be an important effect.},
thus justifying the form \sphereg. At the
surface however, where the boundary conditions are imposed, the 
self consistent form of $k$ must be used. For example Dirichlet ($D$)
and Neumann ($N$) conditions which correspond to reflections,
have $k$ real at the boundary and lead
to $a_l^{D,N} = -(1\mp(-1)^le^{-2ikR})/2$.
More general boundary conditions, mimicking the matching
of internal and external solutions, can be imposed by introducing a
surface impedance.

In the same way as for the point-like object of section 3, 
the bare vertex of the SM equation is given by,
\eqn\vertexg{
V_1({\bf r},{\bf r}') =
{4\pi\over l}
\left( |G({\bf r},{\bf r}')|^2-|G^{(0)}({\bf r},{\bf r}')|^2
\right)
}
Where $G^{(0)}$ is the free green function, given by the same expansion
as \sphereg\ with $a_l = 0$. 
According to \vertex, the charge $Q$ is related to the Fourier transform 
of this vertex at vanishing momenta. This pertubative
result is clearly inappropriate in the interior of the sphere
where $G({\bf r},{\bf r}')$ vanishes, nonetheless corrections from the
interior should not exceed $O(R/l)^3$, and can be neglected in
comparison with the leading term for small spheres.
Using orthogonality properties of the
spherical harmonics, the charge can be written as a sum 
of radial integrals over products of $g_l^{(0)}(r,r')$
and $g_l^{(1)}(r,r')$. These integrals may be simplified using the
asymptotic form of the Bessel functions valid since $kR \gg 1$.
This asymptotic approximation is only good for angular momentum numbers
$l < l_{max}$, where $l_{max} = kR$, which in any case forms a
effective upper limit in the angular momentum sum since the $a_l$'s 
become small for higher $l$ (this approximation is standard in dealing 
with scattering from spheres).
Dropping subleading terms in $1/kl$ we find,
\eqn\chargeg{
Q_1
= {-3\over 4\pi l^2}
\int_R\int_R dr^3dr'^3V_1({\bf r},{\bf r}') 
=
{e^{-2R/l}\over(kl)^2}
\sum_l^{l_{max}} (2l+1)
\left( {a_l+{\bar a_l}\over 2} + |a_l|^2\right)
}
The combination of $a_l$'s is recognisable from the classical
amplitude scattering problem
where a similar sum appears in the absorption cross section.
Using this result, and only keeping the leading term in
$R/l$, we find,
\eqn\chargeabsg{
Q_1
= {3\over 4\pi}
{\sigma_{abs}\over l}
}
Where $\sigma_{abs}$ is the absorption cross section.
It is reassuring that this coincides with the point-like 
result \QandPone. 
In particular, for a black  (totally absorbing) sphere one
has $\sigma_{abs}=\pi R^2$. This implies
$Q_1={3R^2/ 4l}$, a result derived already in \Paas.

For the $D$ and $N$ boundary conditions the charge must vanish
since they correspond to reflection. There is no absorption
and indeed each term in the sum \chargeg\ vanishes when the 
explicit forms of $a^{D,N}_l$ are used.
The calculation of $\sigma_{abs}$ in the case of more general 
boundary conditions (for example a surface impedance) is
a standard problem, the only subtlety arising from the value
of $k$ at the boundary.

The dipole moment can be calculated in a similar way.
We find the general result,
\eqn\dipoleg{
P_1
= {1\over 4\pi l^2}
{\partial\over\partial p}.{\partial\over\partial p'}\tilde V_1|_{0}
=
-(R+l)^2{e^{-2R/l}\over k^2 l}
\sum_l^{l_{max}}
\left( (2l+1){a_l+ \bar a_l\over 2}
 +(l+1)(a_l{\bar a_{l+1}} + {\rm c.c.})\right)
}
The first term in the sum is recognised as the total cross section
$\sigma_{tot} = 2\pi R^2$, but the second term is a combination that never 
appears in the amplitude problem. For boundary conditions based on
a surface impedance we can show that this second term is proportional
to the absorbtion cross section (and therefore vanishes for 
$D$ or $N$ boundary conditions).
Taking only the leading order terms we may write the dipole as,
\eqn\dipoletotg{
P_1
=
{\sigma_{tot}l\over 4\pi }
-{Q_1l^2\over 3}
}
The first term recovers the point object result in the limit $R \ll l$,
whereas the second term is reminiscent of the similar term in 
the formula \Qone\ for a collection of point scatterers.
In case of a small reflecting but non-absorbing
 sphere we have $\sigma_{tot}=2\pi R^2$,
and we find  $P_1=R^2l/2$.

\subsec{Higher order}

We may sum all the higher order corrections to the charge and dipole
moment following the same method as for the point scatterer.
In this case we do not have an explicit form for the Fourier 
transform of the vertex so intermediate formulae are unwieldy.
We start with the form of the vertex for small incoming momentum 
in analogy to formula \Extrascatexp.
\eqn\vertexfts{
V_1({\bf p},{\bf p}')
\buildrel {p \rightarrow 0} \over \longrightarrow
-{4 \pi l^2 \over 3}
Q_1 A^{(R)}_1(p'l,R/l) 
+ {\bf p}.{\bf p}'4\pi l^2 P_1  
A^{(R)}_3(p'l,R/l) 
}
Where we have introduced functions $A^{(R)}_1(p'l,R/l)$
and $A^{(R)}_3(p'l,R/l)$, defined in appendix F. These 
functions become the point scatterer functions $A_1$, $A_3$ in the
limits of small $R$, but are better behaved at large momenta
thus avoiding the problems encountered with the point object.

A calculation of the second order vertex at small incoming
momentum shows that it retains the same form as \vertexfts.
This calculation requires various integrals that are collected
in appendix F.
Proceeding by iteration the full sum can be performed
exactly in the same way as for the point object. 
Our final result takes the same form as \FullQP:
\eqn\FullQPsph{\eqalign{
Q &= {Q_1\over 1 + {4\pi\over 3}Q_1 B_1}
\cr
P &= {P_1\over 1 + 4\pi P_1 B_2}
\cr
}}
Where the integrals over the magnitude of the momentum are
now given in terms of the modified functions $A^R$.
\eqn\FirstQP{\eqalign{
B_1 &= {l^2\over 2\pi^2}\int_0^\infty dq\ q^2 {\cal G}_0(q)
\bigl(A^{(R)}_1(ql,R/l)\bigr)^2\cr
B_2 &= {l^2\over 2\pi^2}\int_0^\infty dq\ q^4 {\cal G}_0(q)
\bigl(A^{(R)}_3(ql,R/l)\bigr)^2\cr
}}
These integrals are well behaved for finite $R/l$ and the major 
contribution comes from the region in which the mesoscopic calculation
is valid. This therefore resolves the difficulty encountered 
with the point scatterer, but the result is only valid for
small $R/l$.



\newsec{Conclusion}

In the mesoscopic regime,
we have shown how to calculate the effective charge and dipole moment
that would be seen in the change of the transmitted light 
due to some simple objects placed inside a diffuse  medium. 
The simplest calculation was for a single point scattering object.
In this case we were able to sum up all higher order terms in the
pertubative expansion (in the strength of the vertex) which
contributed to leading order in the mesoscopic approximation.
However, the point nature of the scatterer was incompatible with the
mesoscopic approximation and this was reflected in the divergent
result.
To resolve this problem we considered extended objects. 
A collection of point scatterers gave finite results, but it
was not possible to analytically sum all the higher order terms.
We therefore turned to a spherical object and 
our main results are the formulae \FullQPsph\ for the charge and 
dipole moment of a small sphere in a multiple scattering medium. 
These results include an infinite sum of higher order corrections
and despite the many simplifying assumptions that went into their
derivation, they should be useful as an independent check of the 
numerical methods in common use.


\bigbreak\bigskip\bigskip\centerline{{\bf Acknowledgements}}\nobreak
This research originated from a pleasant discussion T.M.~Nieuwenhuizen
had with the late Shechao Feng.
We would like to acknowledge useful discussions with J.M.~Luck.

\vfill
\eject
\appendix{A}{Notation}

In this appendix we recall some basic definitions.
The background scatterers that compose the diffuse medium have a 
density $n$ and are individually described
by the t-matrix $t$. The mean free path $l$ is related to the elastic
cross section by $l = 1/n\sigma_{el} = 4\pi/nt\bar t$.
The green function for amplitude propagation in this medium
has its wavevector shifted away from the value 
$k^2 = \omega^2 \epsilon_0$ of the free scalar 
wave equation, $\nabla^2 \psi + k^2\psi = 0$,
to obtain
$ G_0({\bf p})  = (p^2 -k^2 -nt)^{-1} 
= (p^2 -k_R^2 -ik_R/l_{ex})^{-1}$.
In the second form we have explicitly split the t-matrix into
its real and imaginary parts. The real part acts to shift the
frequency, $k_R^2 = k^2  + n \Re e\, t$. This effect
is ordinarily described by an index of refraction,
we shall take this as understood and drop the subscript on $k$. 
The small imaginary part of the t-matrix is most important and
is related by the optical theorem to
the total cross section including both the effect of
elastic scattering and absorption, this is described 
by the extinction mean free path, $n\Im m t /k = 1/l_{ex}$.
Absorption is traditionally described by an 
albedo factor that relates the absorption and elastic 
cross section by $\sigma_{el} = a\sigma_{tot}$ or
$\sigma_{abs} = (1/a - 1)\sigma_{el}$
and also the mean free paths, $l_{ex} = al$.
We shall only allow absorption by the extra, not background,
scatterers.
The optical theorem and relation to the various mean free paths 
and albedo
reads, $1/l = nt\bar t/4\pi = an \Im m\, t/k = a/l_{ex}$.


\appendix{B}{Integrals}

The following pair of integrals, related to the diagrams in
figure 1, appear throughout the work:
\eqn\vertexA{\eqalign{
I_1({\bf q}_1)  &=
\int\! {d^3p\over (2\pi)^3}\ 
G_0({\bf p}+{\bf q}_1) \bar G_0({\bf p})\cr
I_2({\bf q}_1,{\bf q}_2)  &=
\int\! {d^3p\over (2\pi)^3}\ G_0({\bf p+q}_1) G_0({\bf p-q}_2)
\bar G_0({\bf p})\cr}
}
A mesoscopic approximation to these functions is sufficient for
our purposes.

The first integral $I_1$ is the ladder kernel of the SM equation
and may be obtained most conveniently from the mesoscopic 
real space form as the Fourier transform of 
${\cal M}_0 =  e^{-r/l}/4\pi l r^2$,
\eqn\AoneM{
\tilde {\cal M}_0(q) 
= {1\over ql} \int_0^\infty dr\, {\sin qr\over r} e^{-r/l}
= {\tan^{-1}(ql)\over ql}
= A_1(ql)
}
Hence the free green function of the SM equation is
${\cal G}_0(q) = 1/(1-A_1(ql))$.

The integrals \vertexA\ can also be evaluated directly by
performing the integrals over the magnitude of the loop momenta,
and then expanding the angular integrals in
$1/kl$ to find the following mesoscopic limits:
\eqn\vermesoA{\eqalign{
I_1({\bf q}_1)  & \rightarrow
{l\over 4\pi} A_1(q_1l)\cr
I_2({\bf q}_1,{\bf q}_2)  &\rightarrow
{il^2\over 8\pi k} A_2({\bf q}_1,{\bf q}_2)\cr}
}
The properties of the angular integrals $A_1,A_2$ are discussed
in appendix C.


\appendix{C}{Angular Integrals}

The definition and properties of the angular integrals $A_1,A_2$
are given in this appendix.
\eqn\Aone{
A_1(q) =
{1\over 4\pi} \int_{S_2}\! d\hat p \ {1\over 1+ i{\bf q}.\hat{\bf p}}
= {\tan^{-1}(q)\over q}
}
The integral $A_1$ is real and depends only on the length of
the vector ${\bf q}$. Note that $A_1(0) = 1$, and that the 
expansion for small $q$ is $A_1(q) = 1 - q^2/3 + q^4/5 + O(q^6)$.
At large $q$, $A_1$ behaves as $\pi/2q$.

We also require a slightly more complicated integral,
\eqn\Atwo{
A_2({\bf q}_1,{\bf q}_2) =
{1\over 4\pi} \int_{S_2} \! d\hat p \ 
{1\over 1 + i{\bf q}_1.\hat{\bf p}}\ {1\over 1 - i{\bf q}_2.\hat{\bf p}}
}
This expression is again real due to the symmetry 
$\hat{\bf p}\rightarrow -\hat{\bf p}$. It obeys the following,
\eqn\Atwoobey{
A_2({\bf q},{\bf 0})  = A_2({\bf 0},{\bf q}) = A_1(q) 
}

Although a simple expression for this integral in not available,
for the purposes here it is sufficient to consider an expansion 
in the incoming momentum. To leading order in small $q_1$ we have,
\eqn\Atwo{
A_2({\bf q}_1,{\bf q}_2) \approx
A_1(q_2)
-{{\bf q}_1.{\bf q}_2\over q_2^2}\bigl(1 - A_1(q_2)\bigr)
= A_1(q_2) - {\bf q}_1.{\bf q}_2A_3(q_2)
}
where we have introduced $A_3(q) = (1-A_1(q))/q^2$.
If both $q_1$ and $q_2$ are small, 
$A_2({\bf q}_1,{\bf q}_2) \approx 1 - {\bf q}_1.{\bf q}_2/3$.
Some further properties involving angular integrals of $A_2$ are also 
needed,
\eqn\Aangleint{\eqalign{
{1\over 4\pi} \int_{S_2} \! d\hat q \ 
A_2({\bf q},{\bf q}_2) & 
=  A_1(q) A_1(q_2) \cr
{1\over 4\pi} \int_{S_2} \! d\hat q \ 
{\bf q}_1.{\bf q} A_2({\bf q},{\bf q}_2) & 
= {\bf q}_1.{\bf q}_2
q^2 A_3(q) A_3(q_2)\cr
}}
%


\appendix{D}{Matrix Formalism}

Consider each of the 3 diagrams in figure 1 as a separate entry in
a $3\times 3$ matrix. The matrix version of the vertex for small
incoming momentum (3.7) is 
\eqn\matvert{\eqalign{
V_1 &= 
-{4 \pi l^2 \over 3}
{\bf Q}_1 A_1(p'l) 
+ {\bf p}.{\bf p}'4\pi l^2 {\bf P}_1  
A_3(p'l)\cr
{\bf Q}_1 &= {-3\over 4\pi l} 
\left(\matrix{
t\bar t/4\pi &        &             \cr
              & it/2k  &             \cr
              &        & -i\bar t/2k \cr}\right) \cr
{\bf P}_1 &= {l\over 4\pi } 
\left(\matrix{
     0        &        &             \cr
              & it/2k  &             \cr
              &        & -i\bar t/2k \cr}\right) \cr
}
}
This form is preserved at higher order, though $\bf Q$ and $\bf P$
are no longer diagonal, and the measured charge/dipole moment is 
given by the sum of all entries of the respective matrix. 

The advantage of the matrix formalism is in connecting vertices
using a propagator that can take account of the appropriate
combinatorics discussed in the text. To only include ladder
diagrams we write,
\eqn\matprop{
{\bf G} = 
\left(\matrix{
1 & 1 & 1 \cr
1 & 1 & 1 \cr
1 & 1 & 1 \cr}\right)
{\cal G}_0
}
To also include crossed diagrams we have,
\eqn\matpropX{
{\bf G} = 
\left(\matrix{
2 & 2 & 2 \cr
2 & 2 & 1 \cr
2 & 1 & 2 \cr}\right)
{\cal G}_0
}
The higher order diagrams are summed to give,
\eqn\matsum{\eqalign{
{\bf Q} &= 
{\bf Q}_1 \left( 1 + {4\pi\over 3 l}{\bf G Q}_{1}\right)^{-1} \cr
{\bf P} &= 
{\bf P}_1 \left( 1 - {4\pi\over 3l^3}{\bf G P}_1\right)^{-1} \cr
}
}
When evaluated using the ladder propagator we recover the result
(3.12). When the crossed diagrams are included we must explicitly
drop terms that are subleading in the mesoscopic approximation
and are left with the result,
\eqn\matfullQP{\eqalign{
Q &= {Q_1\over 1 + {2\sigma_{el}\over l}(1/a_e-1) B_1}
+Q_1{\sigma_{el}\over l}(1/a_e-1) B_1
= {Q_1\over 1 +{8\pi\over 3}Q_1 B_1}+{4\pi\over 3}Q_1^2 B_1}}
\eqn\matfulp{
P = {P_1\over 1 + {6\sigma_{el} l}(1/a_e) B_2 }+
P_1 3\sigma_{el} l(1/a_e) B_2
= {P_1\over 1 + 8\pi P_1 B_2}+4\pi P_1^2 B_2}
where the added terms correct for the fact that the two-scatterer
diagram has no time-reversed counterpart.
The main change is thus a factor of two in the $B$'s compared with the
ladder result \FullQP.


\appendix{E}{Classical scattering from a sphere}

We collect certain results concerning amplitude scattering from
a sphere \Mierefs. A scattering solution for a field obeying the wave 
equation, with incoming part $e^{ikz}$, and symmetric object, is
\eqn\field{
\phi = \sum_l i^l\sqrt{4\pi(2l+1)}
\left( j_l(kr) + a_l h^{(1)}_l(kr)\right) Y_{l0}(\theta)
}
The coefficients $a_l$ are determined by the boundary conditions.
Besides Dirichlet ($D$) and Neumann ($N$) conditions, a more general
possibility is to assume a surface impedance, $Z$, and require,
$\partial_t \phi = ik \phi = Z\partial_r \phi$, on the surface. These 
coefficients are the same as the ones appearing in the green
function \sphereg.

By calculating the fluxes we find the following cross sections,
\eqn\field{\eqalign{
\sigma_{el} &= {4\pi \over k^2}\sum_l (2l+1) |a_l|^2 \cr
\sigma_{abs} &= {-4\pi \over k^2}\sum_l (2l+1) 
	\left(|a_l|^2 + {a_l + \bar a_l\over 2}\right)\cr
\sigma_{tot} &= \sigma_{el}+\sigma_{abs}\cr
}}
For $D$ and $N$ conditions, $\sigma_{abs} = 0$, and for a large
sphere ($kR \gg 1$), $\sigma_{tot} = 2\pi R^2$.


\appendix{F}{Integrals for Spherical Object}

In the text we have only discussed the position space form
of the bare vertex for the spherical object. 
The following representation in momentum space is
convenient for calculations,
\eqn\vertexftD{\eqalign{
\tilde V_1({\bf p},{\bf p}')
&=
\int d^3 r\int d^3 r' e^{-i{\bf p}.{\bf r}}e^{i{\bf p}'.{\bf r}'}
V_1({\bf r},{\bf r}')\cr
&={4\pi\over l_{ex}}
\int_R dr r^2\int_R dr'r'^2 \sum_{ll'}
\bigl(
\bar a_{l'}g_l^{(0)}\bar g_{l'}^{(1)} + 
a_{l}\bar g_{l'}^{(0)}g_{l}^{(1)} + 
+ a_l\bar a_{l'}g_l^{(1)}\bar g_{l'}^{(1)}\bigr)
\cr
&\quad\times
\int d\hat r\int d\hat r'
 e^{-i{\bf p}.{\bf r}}e^{i{\bf p}'.{\bf r}'}
{(2l+1)\over 4\pi}{(2l'+1)\over 4\pi}
P_l(\cos \gamma)P_{l'}(\cos \gamma)
}}
Where $\gamma$ is the angle between the two vectors $\hat r$ and $\hat r'$.
We have written the mean free path as $l_{ex}$ in this formula to prevent 
confusion with the angular momentum quantum number.
Although not tractable, this form can be used to prove the following
two formulae which are the analogs of the integration formulae \Aangleint\ 
in appendix C.
\eqn\intD{\eqalign{
&\int d^3 r\int d^3 r' a(r) b(r') e^{i{\bf p}'.{\bf r}'}
V_1({\bf r},{\bf r}')\cr
&\quad={4\pi l\over 3} Q_1
\int_R {dr} a(r) e^{(R-r)/l}
\int_R {dr'}{\sin p'r'\over p'r'} b(r')e^{(R-r')/l}\cr
&\int d^3 r\int d^3 r' a(r) b(r') \, i({\bf p}.{\bf r}) e^{i{\bf p}'.{\bf r}'}
V_1({\bf r},{\bf r}')\cr
&\quad=-{\bf p}.{\bf p}'{4\pi l^3\over (R+l)^2} P_1
\int_R {dr r} a(r) e^{(R-r)/l}
\int_R dr'
\left({\sin p'r'\over p'^3r'^2} - {\cos p'r'\over p'^2r'}\right)
b(r') e^{(R-r')/l}\cr
}}
Where $a(r)$ and $b(r)$ are smooth functions. 
The asymptotic form of the Bessel functions has been used and
oscillating terms which contribute at higher order in
in $1/kl$ have been dropped in accordance with the mesoscopic approximation.

The functions $A^{(R)}_1$ and $A^{(R)}_3$ are defined as follows,
\eqn\ARs{\eqalign{
A^{(R)}_1(ql,R/l)
&= \int_R^\infty {dr\over l}
\, {\sin qr\over qr} e^{(R-r)/l}
= {1\over 4\pi}\int_{S_2}\! d\hat p \ 
{e^{-i{\bf q}.\hat{\bf p}R}\over 1+ i{\bf q}.\hat{\bf p}l}
\cr
A^{(R)}_3(ql,R/l)
&= {1 \over q(R+l)}\int_R^\infty {dr\over l}
\, \left({\sin qr\over q^2r^2}-{\cos qr\over qr}\right)
e^{(R-r)/l}\cr
}}
They have the following limits as $R \ll l$,
\eqn\ARlims{\eqalign{
A^{(R)}_1(ql,R/l)
&\rightarrow A_1(ql)\cr
A^{(R)}_3(ql,R/l)
&\rightarrow A_3(ql) = {(1-A_1(ql))\over (ql)^2}\cr
}}
%


\listrefs
\bye